\documentclass{article}
\usepackage{ijcai15}

\usepackage{times}
\usepackage{amsfonts}

\usepackage{latexsym} 
\usepackage{tikz}
\usepackage{amssymb,amsmath}
\usepackage{xspace}

        \newtheorem{lemma}{Lemma}%
        \newtheorem{observation}{Observation}%
        
                \usepackage{colortbl,color,array,hhline}
        \definecolor{green}{RGB}{0, 102, 0}

\newcommand{\PNE}{pure Nash equilibrium\xspace}
\newcommand{\PNEs}{pure Nash equilibria\xspace}
\newcommand{\nPNE}{simple pure Nash equilibrium\xspace}
\newcommand{\nPNEs}{simple pure Nash equilibria\xspace}

\title{Online Fair Division: analysing a Food Bank problem}
\author{Martin Aleksandrov, Haris Aziz, Serge Gaspers \\UNSW and NICTA, Kensington 2033, Australia
\AND Toby Walsh
\\UNSW and NICTA, Kensington 2033, Australia}

\begin{document}

\maketitle

\begin{abstract}
We study an online model of fair division
designed to capture features of a real world
charity problem. We consider two simple mechanisms
for this model in which agents simply declare what
items they like. We analyse several axiomatic
properties of these mechanisms 
like strategy-proofness and envy freeness.
Finally, we perform a competitive
analysis and compute the price of anarchy.
\end{abstract}

\section{Introduction}

Resource allocation is a fundamental problem
facing society. How do we share scarce and often costly
resources between different parties?
Due to environmental, economic and technological
changes, there is an every increasing pressure
on the allocation of resources. The theoretical 
foundations of resource allocation have been developed 
using simple abstract models. For example,
one simple model for resource allocation is 
fair division. Fair division problems are typically 
categorised along several orthogonal dimensions: 
divisible or indivisible goods, centralised or decentralised 
mechanisms, cardinal or ordinal preferences, etc.\ 
(e.g.\ \cite{rasurvey}). However, 
such categories do not capture the richness of many
real world fair division problems. 
This has motivated a call to develop more
complex and realistic models and mechanisms 
\cite{waaai15}. In this paper, we respond to this call 
by studying mechanisms for an online
fair division problem first proposed in
\cite{wki14}. 

\section{The Food Bank problem}

Unfortunately, even in developed countries, poverty
remains a serious problem. For example,
the 2012 report ``{\em Poverty In Australia}''
estimated that over 2 million people 
(12.5\% of the population) are within
the official definition of poverty (less
than half the median income) \cite{pia}. 
Amongst the young and old, the
statistics are even worse (roughly 1 in
6 children, and 1 in 4 pensioners). 
These people struggle to feed themselves
and increasingly call upon food banks to
help. Food Bank Australia sees the
demand on their services increase by
over 10\% per annum. For this reason, they
are keen to improve the efficiency of their
operations.

In cooperation with a social startup, FoodBank Local,
we have been helping Food Bank Australia develop
technologies to operate more effectively. So far,
this has involved building an app to help collect
and deliver donated food. This app uses our vehicle routing
solver to route their trucks. We are
now turning our attention to how the donated
food is allocated to different charities. 
This is an interesting fair division problem.
It has many traditional features. 
We want to allocate food {\em fairly} between the
different charities that feed different
sectors of the community. Goods are mostly {\em indivisible}. 
The allocation does not use money as these are all charities. 
However, the problem also has other features not traditionally found
in the academic literature on fair division. One of the main
novelties is that it is {\em online}. Food is donated
throughout the day and we must start allocating
and distributing it almost immediately, before we know what
else will be donated. We have therefore 
formulated an online model of their fair division problem, 
and studied mechanisms that can
fairly and efficiently allocate the 
donated food. 

\section{Online fair division}

\newtheorem{mytheorem}{Theorem}
\newtheorem{mydefinition}{Definition}
\newtheorem{myexample}{Example}
\newcommand{\myproof}{\noindent {\bf Proof.\ \ }}
\newcommand{\myproofsketch}{\noindent {\bf Proof sketch.\ \ }}
\newcommand{\myqed}{\mbox{$\Box$}}
\newcommand{\myOmit}[1]{}
\newcommand{\myblacksquare}{$\blacksquare$}
\newcommand{\mylike}{\mbox{\sc Like}}
\newcommand{\myblike}{\mbox{\sc Balanced Like}}

We have $k$ agents. Each agent has
some (private) utility for the $m$ items. 
One of the $m$ items appears at each time step,
and the allocation mechanism must
assign it to one of the agents. 
The next item is then revealed. 
This continues for $m$ steps. 
To allocate items in this online model, 
we consider a simple class of bidding
mechanisms in which agents merely
declare if they like items or
not. For instance, the \mylike\ mechanism
allocates the next item uniformly at random
between agents that declare that they like the item. 
An allocation is a possible outcome of the
\mylike\ mechanism if each item
is given to an agent that values it,
whilst an allocation is the necessary outcome
if no two agents like the same item, and each 
item is given to the agent that values it, or
to no one if no agent likes it. 

One problem with the \mylike\ mechanism
is that agents can get unlucky. It is possible
for them to bid for every item but have
every coin toss go against them and not be 
allocated anything at all. This is highly
undesirable in our Food Bank setting. A 
whole sector of the population will then 
not be fed that night. 
We therefore consider a slightly more sophisticated mechanism
that helps tackle this problem. 
The \myblike\ mechanism tries to balance
the number of items allocated to agents
compared to the \mylike\ mechanism.
It allocates
the next item uniformly at random between 
those agents that value it that have
so far received the fewest items. 
The \myblike\
mechanism is less likely to leave agents empty 
handed than the \mylike\ mechanism. 
In particular, an agent is {\em guaranteed}
to be allocated at least one item for every $k$ items
that they like. However, there is no 
guarantee that it {\em necessarily} returns
balanced allocations. 


Given the order of items, we can 
compute the actual outcome of both the
\mylike\ and \myblike\ mechanisms
efficiently. Each 
of the $m$ steps takes $O(k)$ time.
Supposing agents bid sincerely,
computing the probability an agent
gets a particular item, as well as
their expected utility is more 
challenging as there are $O(k^m)$ possible
outcomes. 
With the \mylike\ mechanism,
the probability that agent $i$
gets item $j$ is 
simply $1/q_j$ where $q_j$ is the
number of agents who like $j$.
The expected utility is then
$\sum_{j=1}^m \frac{u_i(j)}{q_j}$
where $u_i(j)$ is the private utility of 
agent $i$ for item $j$. 
With the \myblike\ mechanism,
we can compute the probability that an agent
gets a particular item 
using dynamic programming.
This exploits the fact that the
mechanism is Markovian. It doesn't 
care how we get to a particular state,
just how many items each agent has at
this point. The states represent the
number of items allocated to each
agent. 
\myOmit{We start in state $(0,\ldots,0)$,
which represents the state in which 
no agents have any items. We record in 
$(0,\ldots,0)$ the probability of being in this state.
Hence, we record in $(0,\ldots,0)$
the value 1. The update rule then 
shares the probability recorded in
a given state amongst those states 
to which we can transition.
For example, if the first item
is liked by just the first and second agent,
we record $\frac{1}{2}$ of the 1 in
the state $(1,0,\ldots,0)$
and the other 
$\frac{1}{2}$ of the 1 in
the state $(0,1,\ldots,0)$.
Using this dynamic program,}
We can compute the 
probability that an agent gets
a particular item, as well as
the expected utility of an agent
in $O(m^k)$ space and time. 
Note that
$k$ is typically smaller than $m$
so $O(m^k)$ is likely better than
$O(k^m)$. 

\section{Strategy-proofness}

As is common in the literature, we will consider the
axiomatic properties of 
these mechanisms. For example,
we say that a mechanism for online
fair division is {\em strategy-proof}
if and only if, with knowledge
of the items still to be revealed, the order in which
they will be revealed, and the
private utilities of the other agents, 
an agent cannot increase
their expected utility by bidding 
differently to their true preferences.
We might prefer strategy-proof mechanisms
as agents cannot manipulate the outcome
to improve their utility at the expense
of agents who are less sophisticated or
knowledgeable. 

\begin{mytheorem}\label{thm:LikeSP}
The \mylike\ mechanism is strategy-proof. 
\end{mytheorem}
\myOmit{\myproof
First, an agent has no
incentive to bid for an item that they
do not value. This will not increase 
their expected utility. Second, they will 
always bid for items that they value
as not bidding only decreases their expected
utility. An agent's best action is thus
to bid sincerely. 
\myqed
}

With the \myblike\ mechanism, 
balancing the size of
the allocations has an unfortunate 
side effect: an agent can now manipulate the 
outcome to increase their expected utility
by bidding strategically.
In particular, an agent may choose not to bid
for an item now in the knowledge that this will
bias future allocation rounds in their favour. 
Such manipulations may decrease
the equitability of the final allocation. 

\begin{mytheorem}\label{thm:BalancedLikeSP}
The \myblike\ mechanism is not strategy-proof
even when restricted to 0/1 utilities. 
\end{mytheorem}
\myproof
Suppose we are allocating the items $a$, $b$ and $c$
in this order between agents $1$, $2$ and $3$ with agent 1
having utility 1 for all items,
agent 2 for $a$ and $c$,
and agent 3 for $b$ alone. 
Then bidding sincerely
gives agent 1 an expected utility
of $\frac{9}{8}$ but this can
be increased to $\frac{5}{4}$ if
agent 1 strategically bids only
for items $b$ and $c$ supposing the 
other agents bid sincerely. 
\myOmit{
with the following utilities.

\begin{tabular}{r c c c}
        & $a$ & $b$ & $c$\\ \hline
Agent 1 & $1$ & $1$ & $1$\\
Agent 2 & $1$ & $0$ & $1$\\
Agent 3 & $0$ & $1$ & $0$
\end{tabular}

\noindent
The items are allocated in alphabetical order. Figure 1 depicts all possible allocations of $a$, $b$ and $c$.

\begin{figure}[h]
\begin{center}
\begin{tikzpicture}[xscale=1,yscale=0.5]
 \tikzstyle{ag1}=[very thick];
 \tikzstyle{ag2}=[very thick, dash pattern=on 5pt off 2pt];
 \tikzstyle{ag3}=[very thick, dotted];
 
		\node at (0,1.875) (A) {$a$};
		
		\node at (1.5,3) (B) {$b$};
     	\node at (1.5,0.75) (C) {$b$};
     	
     	\node at (3,3) (D) {$c$};
    	\node at (3,1.5) (E) {$c$};
       	\node at (3,0) (F) {$c$};
    	    
    	\node at (4.5,3) (G) {};
      	\node at (4.5,2) (H) {};
     	\node at (4.5,1) (I) {};
     	\node at (4.5,0) (J) {};
     	
     	\draw[ag1] (A)--(B) (C)--(E)--(H) (F)--(J);
     	\draw[ag2] (A)--(C) (D)--(G) (E)--(I);
     	\draw[ag3] (B)--(D) (C)--(F); 
\end{tikzpicture}
\end{center}
\caption{agent 1-solid, agent 2-dashed, agent 3-dotted}
\end{figure}

\noindent
Suppose agents act sincerely and bid for all items that they like.
Then there is one outcome where agent 1 receives two items, namely
when agents receive items in the order $(2,1,1)$.
This happens with probability $\frac{1}{8}$. 
In all other outcomes, agent 1 is 
allocated exactly one item. Hence the expected
utility of agent 1 is $\frac{1}{8}\cdot 2 + \frac{7}{8}\cdot 1 = \frac{9}{8}$. 

Suppose agent 1 acts strategically and bids only
for items $b$ and $c$, while the other two agents
bid sincerely. His strategic move now increases the probability 
of the outcome where he receives two items to $\frac{1}{4}$
and his expected utility becomes $\frac{1}{4}\cdot 2 + \frac{3}{4}\cdot 1 = \frac{5}{4} > \frac{9}{8}$. 
%
%
}{
\myqed

It is a strong assumption to suppose that a
strategic agent has full knowledge of the 
items still to be revealed, the order in which
they will be revealed, and the private utilities
of the agents for these items.
In practice, agents may only have partial
knowledge. This will greatly limit 
the willingness of, say, a risk averse agent
to be strategic. For instance, if there is
a chance that only items that you do not value
will arrive in the future, a risk averse agent will always
sincerely bid for an item that arrives now which
they value. Interestingly,
when limited to just two agents and bivalent
utilities, the \myblike\ mechanism
becomes strategy-proof even under our strong
assumption of complete knowledge. 

\begin{mytheorem}
With only 2 agents and 0/1 utilities,
the \myblike\ mechanism is strategy-proof. 
\end{mytheorem}
\myproof (Sketch)
Without loss of generality, it is sufficient to prove that truth-telling is the dominant strategy for agent $1$. The general idea of the proof is that we focus on the last item that agent $1$ misreported. We show that agent $1$ does at least as well or strictly better by expressing a preference in which he does not misreport about this item. By induction, agent $1$ does not have an incentive to misreport any item.

Consider any ordering of the items where in each round $i$, item $o_i$ is allocated to either to agent $1$, $2$, or neither of them. We want to show that agent $1$ has no incentive to be untruthful even if he is aware of the ordering beforehand. 
We view the allocation process as an allocation tree as follows.
A node labelled $(i,(x,y))$ denotes a decision point in the allocation process when the $i$-th item is allocated, $x$ denotes the number of items already allocated to agent $1$ and $y$ denotes the number of items already allocated to agent $2$. Depending on what allocation decision is taken from node $(i,(x,y))$, we arrive at child node of $(i,(x,y))$,
which is $(i+1,(x+1,y))$, $(i+1,(x,y+1))$, or $(i+1,(x,y))$, depending on whether item $o_i$ is allocated
to agent 1, agent 2, or neither of them.
%
Let $T(i, (x,y))$ be the allocation sub-tree tree starting from a node at round $i$ in which $x$ items have been allocated to agent $1$ and $y$ items have been allocated to agent $2$. Let $U_1(T(i, (x, y)))$ be the total expected utility for agent $1$ starting from the node $(i, (x, y))$ when agents report truthfully.

\begin{observation}\label{obs:difference}
The allocation tree has the following memory-less property: 
if there is a node $v$ labelled $(i,(x,y))$ and a node $v'$ labelled $(i,(x',y'))$
such that $x-y = x'-y'$, then the sub-trees rooted at $v$ and $v'$ are identical,
irrespective of how items were allocated previously.
%
\end{observation}

The following lemmas can be proved by analysing the allocation trees.
The base cases $i=m$ are trivial. For the induction, we prove that if the statement holds for $i+1$ to $m$, then it also holds for $i$.

\begin{lemma}\label{lemma:-1+1}
For any integers $x$ and $y$, and for all $i=1,\ldots, m$,
	$U_1(T(i, (x, y)))\geq U_1(T(i, (x-1,y+1)))$
\end{lemma}

\begin{lemma}\label{lemma:-1}
For any integers $x$ and $y$, and for all $i=1,\ldots, m$,
	$U_1(T(i, (x, y)))\geq U_1(T(i, (x-1,y)))$
\end{lemma}

\begin{lemma}\label{lemma:y+1}
For any integers $x$ and $y$, and for all $i=1,\ldots, m$,
	$U_1(T(i, (x, y)))\geq U_1(T(i, (x,y+1)))$
\end{lemma}

Using these lemmas, we can prove that agent 1 has no incentive to misreport any item.
Let $u_1'$ denote agent 1's insincere bid, let $o_i$ denote the last item for which agent 1 does not bid
sincerely, and let $u_1$ denote the bid obtained from $u_1'$ by voting sincerely for item $o_i$.
Let us further suppose that we are at node $(i,(x,y))$ and $u_1'(o_i)=1$ whilst $u_1(o_i)=0$.
By Observation~\ref{obs:difference} and Lemma~\ref{lemma:y+1}, there is no incentive for agent $1$ to approve an unapproved item. Therefore, assume that agent 1 does not
bid for an item he likes: $u_1'(o_i)=0$ and $u_1(o_i)=1$.
%
%
We focus on the node $(i,(x,y))$ which leads to different subtrees depending on whether agent $1$ reports $u_1$ or $u_1'$.

\begin{itemize} \itemsep=0pt
	\item If agent $2$ does not bid for $o_i$, then agent $1$ gets $o_1$ for sure under $u_1$ but no one gets $o_i$ if agent $1$ reports $u_1'$.
Under $u_1$ we arrive at a node labelled $(i+1,(x+1,y))$, whereas under $u_1'$ we arrive at a node labelled $(i+1,(x,y))$. By Lemma~\ref{lemma:-1}, $u_1$ yields at least as much utility as $u_1'$ since $U_1(T(i+1,(x+1,y)))\geq U_1(T(i+1,(x,y)))$.

\item If agent $2$ bids for $o_i$ and $x<y$, then agent 1 receives item $i$ under $u_1$
and agent 2 receives the item under $u_1'$. Since, by Lemma~\ref{lemma:-1+1}, $U_1(T(i+1, (x+1, y)))\geq U_1(T(i, (x,y+1)))$, reporting
$u_1$ yields at least as much utility to agent 1 as $u_1'$. 

\item If agent $2$ bids for $o_i$ and $x=y$, then under $u_1$ there are two children labeled $(i+1,(x+1,y))$ and $(i+1,(x,y+1))$ of $(i,(x,y))$. But under $u_1'$, there is only one child  $(i+1,(x,y+1))$ of $(i,(x,y))$.  By Lemma~\ref{lemma:-1+1}, $U_1(T(i+1, (x+1, y)))\geq U_1(T(i+1, (x,y+1)))$ and thus
$u_1$ yields at least as much utility to agent 1 as $u_1'$.

\item Finally, if agent $2$ bids for $o_i$ and $x>y$, then agent 2 receives item $o_i$, no matter how agent $1$ bids. Therefore, agent $1$ has no incentive to report $u_1'$ rather than $u_1$.
\end{itemize}
This completes the proof of the theorem.
\myqed

Intuitively, one might hope that this theorem can be generalized to an arbitrary number of agents where each item
is valued by at most 2 agents. However, the example in the proof of Theorem~\ref{thm:BalancedLikeSP} shows that this is not possible, even for 3 agents.
It is also easy to give examples
with more general utilities where 
the \myblike\ mechanism is not
strategy-proof even with 2 agents.

\begin{myexample}
Consider 2 agents and 2 items, $a$ and $b$. 
Agent 1 has utility $\frac{1}{2}$ for both 
items, and agent 2 has utility $\frac{1}{4}$
for item $a$ and $\frac{3}{4}$
for item $b$, normalized to sum up to 1.
If agents bid sincerely for both items, 
then agent 2 has an expected utility of
$\frac{1}{2}$. However, by bidding strategically
only for item $b$, agent 2 can increase
their expected utility to $\frac{3}{4}$. 
Agent 1  has no incentive
to bid strategically and receives the optimal utility
of $\frac{1}{2}$ in both cases. In this case,
strategic behaviour leaves the egalitarian
welfare unchanged, and actually increases the overall
utilitarian welfare. 
\end{myexample}

%

\section{Impact on welfare}
\label{sec:welfare}

Strategic play can have both a positive or
negative effect on welfare. We consider \PNEs in 
which no agent can get strictly greater expected
utility by changing their strategy. 
Although the \mylike\ mechanism is strategy-proof,
there are \PNEs that have much smaller egalitarian and utilitarian
welfare than sincere play for both
mechanisms.

\begin{mytheorem}
	There are instances with 0/1 utilities and $k$ agents, where the egalitarian and
	utilitarian welfare of sincere play in the \mylike\ and \myblike\ is $k$ times the corresponding welfare of at least one \PNE. 
\end{mytheorem}
\myproof
 Consider an instance with $k$ agents and $k$ items.
 For each $i\in\{1,\dots,k\}$, agent $i$ values item $i$ and no other item.
 For sincere play, item $i$ is assigned to agent $i$ in both the \mylike\ and \myblike\ mechanisms, giving an egalitarian utility
 of $1$ and a utilitarian utility of $k$.
 Let us now consider the \PNE where each agent bids for all items.
 In the \mylike\ mechanism,
 with these bids, each agent is allocated each item with probability $1/k$.
 Since each agent values exactly one item, this gives an expected egalitarian welfare
 of $1/k$ and an expected utilitarian welfare of $1$.
 In the \myblike\ mechanism, each agent is allocated exactly one item.
 The probability that this item is the one she likes is $1/k$, giving
 again an expected egalitarian welfare
 of $1/k$ and an expected utilitarian welfare of $1$.
\myqed

In the \mylike\ mechanism, a \PNE cannot lead to greater egalitarian
or utilitarian welfare than sincere play as no player has an incentive to not bid
for an item she likes.
The example in the last proof has
many agents that bid for items for which they
have no value. Such bids do not hurt an individual's (expected)
utility but neither do they help. 
We will consider a subset of \PNEs
by supposing a small utility cost to liking (or taking delivery of) an item.
We call these \emph{simple} \PNEs.
Note that sincere play is the only \nPNE for the
\mylike\ mechanism, and therefore, there is no difference in welfare between
sincere play and \nPNEs.

For the \myblike\ mechanism, \nPNEs have
the same utilitarian utility as sincere play, as each item is assigned to
an agent who likes it. However, we will show that a \nPNE
may have smaller or greater egalitarian utility than sincere play.

\begin{mytheorem}
	There are instances with 0/1 utilities where the expected egalitarian welfare of sincere play in the \myblike\ mechanism is strictly greater than the expected egalitarian welfare of each \nPNE.
\end{mytheorem}
\myproof
 The proof of Theorem~\ref{thm:BalancedLikeSP} gives an instance where the unique \nPNE
 has less expected egalitarian utility than sincere play.
\myqed

\begin{mytheorem}
	There are instances with 0/1 utilities where the expected egalitarian welfare of sincere play in the \myblike\ mechanism is strictly smaller than the expected egalitarian welfare of each \nPNE.
\end{mytheorem}
\myproof
 Consider the following instance.
 
 \begin{tabular}{r c c c c c c}
 	& $a$ & $b$ & $c$ & $d$ & $e$ & $f$\\ \hline
 	Agent 1 & $1$ & $1$ & $1$ & $0$ & $0$ & $0$\\
 	Agent 2 & $1$ & $0$ & $1$ & $0$ & $1$ & $1$\\
 	Agent 3 & $1$ & $1$ & $0$ & $1$ & $0$ & $1$
 \end{tabular}
 
 \noindent
 Running the \myblike\ mechanism, one always obtains an allocation with egalitarian welfare 1, except when the items are allocated to the agents according to the sequence of agents $(2, 1, 1, 3, 2, 3)$, in which case the egalitarian welfare is 2.
 By analysing the allocation tree of the \myblike\ mechanism, one can see that the
 instance has a unique \nPNE, which favours this allocation.
 
 \begin{tabular}{r c c c c c c}
 	& $a$ & $b$ & $c$ & $d$ & $e$ & $f$\\ \hline
 	Agent 1 & $0$ & $1$ & $1$ & $0$ & $0$ & $0$\\
 	Agent 2 & $1$ & $0$ & $1$ & $0$ & $1$ & $1$\\
 	Agent 3 & $1$ & $1$ & $0$ & $1$ & $0$ & $1$
 \end{tabular}
 
 \noindent
 We obtain an expected utilitarian utility of $13/12$ for sincere play and $9/8$ for the \nPNE. 
\myqed

\section{Fairness}

How fair are these mechanisms? Is the
\myblike\ mechanism more fair
in some sense than the \mylike\ mechanism.
Since
the outcome of our mechanisms are random, we consider 
fairness notions 
both ex post (with respect to the actual allocation achieved in
a particular world) and ex ante (with respect to the expected
utility over all possible worlds). 
\myOmit{
There are various notions of fairness
studied in the fair division literature (envy-freeness,
proportionality, \ldots).

\subsection{Envy freeness}
}
One notion of fairness commonly considered
in the fair division literature is envy freeness \cite{cakecut}. 
An agent {\em envies ex post} another
agent if their utility of the other agent's allocation is 
greater than the utility of their allocation. 
Similarly, an agent {\em envies ex ante} another
agent if their expected utility of the other agent's allocation is 
greater than their expected utility of their allocation. 
A mechanism is {\em envy free ex post/ex ante} if
no agent envies another ex post/ex ante. 
We also consider a weaker notion.
An agent has {\em bounded envy ex post} of another
agent if there exists a constant $r$
such that in every case 
their utility of the other agent's allocation is 
at most $r$ greater than their utility of their allocation.
Similarly, 
an agent has {\em bounded envy ex ante} of another
agent if there exists a constant $r$
such that 
their expected utility of the other agent's allocation is 
at most $r$ greater than their expected utility of their allocation.
We say that 
a mechanism is {\em bounded envy free ex post/ex ante} if
each agent has bounded envy ex post/ex ante of every other agent.

If a mechanism is envy free ex post/ex ante then it
is bounded envy free ex post/ex ante, whilst
if a mechanism is (bounded) envy free ex post
then it is (bounded) envy free ex ante. 
It is easy to show that no mechanism for indivisible
items that allocates all items can be envy free ex post:
suppose we have one indivisible item and two
or more agents who value it. 
Regarding the other envy free properties, we prove the following results. 

\begin{mytheorem}
Supposing agents act sincerely 
then the \mylike\ mechanism is envy free ex ante. It is 
not bounded envy free ex post, even with 0/1 utilities
and 2 agents. 
\end{mytheorem}
\myproof 
To prove envy freeness
ex ante, we perform induction over the number of
items. In the base case, 
we have no items to allocate, 
each agent receives 
an expected utility of 0, and no agent
envies another ex ante. 
For the induction step, 
we suppose the allocation of the first $m-1$ items is
envy free ex ante,
and consider the $m$th item which is allocated. 
Suppose $j$ ($\leq k$)
agents have non-zero utility for the $m$th item.
Then each agent receives 
this item in $\frac{1}{j}$ of the possible
worlds. This means that the new allocation
remains envy free ex ante. 

To show that the \mylike\ mechanism is not bounded
envy free ex post even with 0/1 utilities, 
suppose $2$ agents have utility 1 for all $m$ items. There is
one outcome in which the first agent gets
lucky and is assigned every item. However,
in this case, the other agent assigns a utility
$m$ greater to the first agent's allocation
than to their own (empty) allocation.
\myqed

As the \mylike\ mechanism is strategy-proof, it seems 
reasonable to suppose agents act sincerely.
By comparison, when limited to 0/1 utilities,
the \myblike\ mechanism
is both envy free ex ante, 
and bounded envy free ex post. 

\begin{mytheorem}
Supposing agents act sincerely and all utilities
are 0 or 1, the \myblike\ mechanism is envy free ex ante
and bounded envy free ex post. 
\label{thm-bl-ef}
\end{mytheorem}
\myproofsketch
Both proofs use induction on the number of
items. For envy freeness ex ante, 
the induction step uses case analysis
to show that the expected increase in utility
for an agent is at least as large as their
expected increase in utility for the allocation 
of any other agent. 
For bounded envy freeness ex post, 
the induction step again uses case analysis
to show that the envy is at most 1 unit. 
\myOmit{
To show that the
mechanism is envy free ex ante, we 
use induction on the number of items. 
In the base case of
the induction, we have no items to allocate, 
and the induction hypothesis holds trivially. 
For the inductive step, 
we suppose the allocation of the first $m$ items 
is envy free ex ante, 
and consider the $m+1$th item which is allocated. 
We have some probability distribution over
allocations of the first $m$ items. Consider any allocation
of these $m$ items
with non-zero probability. There are three cases to consider.
In the first case, an agent (let's call them agent 1)
likes the item, 
and has a probability of $\frac{1}{j}$ for some $j \leq k$ of
being allocated the item conditional on the
allocation of the first $m$ items. 
In the second case, an agent (let's again call them agent 1)
likes the item, 
and has a probability of 0 of
being allocated the item conditional again
on the allocation of the first $m$ items. 
In the third case, an agent (let's call them agent 2)
does not like the item, and is not allocated the item. 
In the first case, 
the expected utility of agent 1 for their allocation
increases by $\frac{1}{j}$ multiplied by the probability
of this allocation of the first $m$ items. This exactly 
matches the increase in agent 1's expected utility
of the allocations of the other $j-1$ agents
who have a non-zero chance of being allocated the $m+1$th item.
Note that agent 1's expected utility
of the allocations of any other agent
is unchanged. As agent 1 was envy free ex ante
for the first $m$ items (induction hypothesis), 
they remain envy free ex ante over all $m+1$ items.
In the second case, 
agent 1 must have been allocated more items from the first
$m$ items than the $j$ agents who 
have a non-zero probability of being allocated
the $m+1$th item conditional on the allocation
of the first $m$ items. The only agents that agent 1
might now envy are one of these $j$ agents.
However, as agent 1 is allocated at least as many items
as each of these agents in the allocation of the $m+1$ items, 
the expected utility of agent 1 
for one of the allocations of these $j$ agents will
increase but cannot be greater than agent 1's expected utility of
their own allocation. 
In the third case, 
agent 2 does not like the item, and is not allocated the item. 
Hence, this case does not contribute to
any change in the expected utility agent 2 has
for their or any other agent's allocation. 
Thus, the allocation of the $m+1$ items 
is envy free ex ante. 

To show that the \myblike\ mechanism
is bounded envy free ex post, we 
use induction on the number of items to 
prove that an agent's utility of another
agent's allocation is at most one unit of 
utility greater than their utility of their own
allocation. In the base case of
the induction, we have no items to allocate, 
and the induction hypothesis holds trivially. 
For the inductive step, 
we suppose the allocation of the first $m$ items has an
envy ex post bounded by at most one unit of utility, 
and consider the $m+1$th item which is allocated. 
Without loss of generality,
we suppose that the item is allocated to agent 1
(who assigns 1 to the utility of this item). 
Clearly, agent 1
does not envy the other agents more than before the $m+1$th item was allocated
(as being allocated the last item
only increases their total utility). 
Consider some other agent, say 2.
Since
agent 2 had bounded envy of all other agents over the first $m$ items
(by the induction hypothesis), the only agent that agent 2 could
now envy is agent 1. 
There are two cases. In the first case, agent 2 likes
the $m+1$th item. There are two subcases to this first case.
In the first subcase, agent 2 was allocated the same number
of items as agent 1 from the first $m$ items (but
was unlucky and did not win the random draw for the $m+1$th item). 
But as agent 1 now has at most
one more item than agent 2, this envy is limited to
at most one unit of utility. 
In the second subcase, agent 2 was allocated more
items than agent 1 from the first $m$ items (and so did not
enter the random draw for the $m+1$th item). 
But as agent 1 
now has at most the same number of items 
as agent 2, agent 2 cannot assign greater utility to
agent 1's allocation than their own. 
There remains the second case in which agent 2 does not like
the $m+1$th item which is allocated to agent 1. In this case, 
agent 2 remains with bounded envy in the allocation of $m+1$ items
since they had bounded envy of the allocation of the first $m$ items,
and their utility of the other allocations of the 
other agents remains unchanged when agent 1 is allocated
the $m+1$th item.
}
\myqed

It is not hard to show that with general utilities,
the \myblike\ mechanism
is no longer envy free ex ante, 
or bounded envy free ex post (or even, ex ante).
Balancing the allocation of items may prevent an agent who
values an item greatly from being allocated it. 

\begin{myexample}
Consider 2 agents and 2 items, $a$ and $b$. Suppose
agent 1 has utility 0 for $a$ and $p$ for $b$,
but agent 2 has utility $1$
for item $a$ and $p-1$
for item $b$ where $p>2$. Note that both agents
have the same sum of utilities for the
two items. If agents bid sincerely 
then agent 2 gets an expected utility of just
$1$ and envies ex ante agent 1's allocation
which gives agent 2 an expected utility of $p-1$.
As $p$ is unbounded, agent 2 does not have
bounded envy ex post or ex ante of agent 1. 
\end{myexample}

To conclude, on the basis of envy freeness, provided
utilities are 0/1 (or close to this), we 
might consider the \myblike\ mechanism
to be somewhat more fair than the \mylike\ mechanism.
On the other hand, 
when utilities are not 0/1 (or close to this), we 
might consider the \myblike\ mechanism
to be somewhat less fair than the \mylike\ mechanism.

\myOmit{
\subsection{Proportionality}

Another notion of fairness is 
proportionality. 
A mechanism is {\em proportional ex post} if
each of the $k$ agents 
receives at least $\frac{1}{k}$ of their
total utility in any possible allocation. 
A mechanism is {\em proportional ex ante} if
each agent receives in expectation
at least $\frac{1}{k}$ of their
total utility. 
It is easy to show that no mechanism for indivisible
items can be proportional ex post.
Suppose we have one indivisible item and two or more agents that
value the item. It is, however, possible
to be proportional ex ante. 

\begin{mytheorem}
Supposing agents act sincerely, 
then the \mylike\ mechanism is proportional ex ante.
\end{mytheorem}
\myproof 
Immediate.
\myqed

In comparison, the \myblike\ mechanism
is not proportional ex ante in general, even with just 2 agents.
Balancing the allocation of items may prevent agents 
being allocated items that they value  greatly. 

\begin{myexample}
Consider 2 agents and 2 items, $a$ and $b$. Suppose
agent 1 has utility 0 for $a$ and 1 for $b$,
but agent 2 has utility $\frac{1}{4}$
for item $a$ and $\frac{3}{4}$
for item $b$. Note that the utilities of
both agents are normalized to sum to 1.
If agents bid sincerely 
then agent 2 has an expected utility of just
$\frac{1}{4}$. 
\end{myexample}

However, when utilities are 0/1, the
\myblike\ mechanism becomes proportional ex ante. 

\begin{mytheorem}
Supposing agents act sincerely and 
utilities are 0 or 1, then the \myblike\ mechanism
is proportional ex ante. 
\end{mytheorem}
\myproof 
Immediate from Theorem \ref{thm-bl-ef} and
the fact that envy freeness implies proportionality.
\myqed

Thus, on the basis of proportionality, we 
might consider the \myblike\ mechanism
to be somewhat less fair than the \mylike\ mechanism.
}

\section{Competitive analysis}

A powerful technique to study online mechanisms
is competitive analysis \cite{competitiveanalysis}.
This identifies the loss in efficiency due to
the data arriving in an online fashion. 
We say that a randomized mechanism $M$ for online
fair division is 
$c$-competitive from an egalitarian/utilitarian perspective
iff there exists a constant $a$ such that 
whatever the input sequence of items $\pi$:
$$ SW_{OPT}(\pi) \leq c \cdot E[SW_{M}(\pi)] + a$$
where $E[SW_M(\pi)]$ is the expected egalitarian/utilitarian
social welfare of the mechanism on $\pi$,
and $SW_{OPT}(\pi)$ is the optimal egalitarian/utilitarian
social welfare of an (offline) assignment. We suppose
agents bid sincerely. In the next section,
we consider the price of anarchy, which is 
essentially the competitive ratio when agents bid 
strategically. The following results
hold irrespective of the model of the adversary
(oblivious, or adaptive offline). 

The \mylike\ mechanism is competitive
when the number of agents is bounded,
even with general utilities. 

\begin{mytheorem}
With $k$ agents, the \mylike\
mechanism is $k$-competitive from an egalitarian or
utilitarian
perspective. 
\end{mytheorem}
\myproof
With the \mylike\ mechanism, the worst case for every agent is 
that every other agent bids against
them. Hence, the worst case is that their
expected social welfare 
is $\frac{1}{k}$ the smallest sum of utilities. 
By comparison, the best case for an agent 
is that they receive the sum of their utilities. Hence, the
competitive ratio from an egalitarian
or utilitarian perspective is at worst $k$.
From an egalitarian perspective, 
this bound is met even when utilities are just 0 or 1.
Consider $k^2$ items being
divided between $k$ agents. The first agent has utility of
1 for the first $k$ items and $0$ for all remaining items.
The other agents have utility 1 for all items. 
The optimal offline allocation achieves egalitarian
social welfare of $k$ units, but 
expected 
egalitarian social welfare of the \mylike\
mechanism is just 1 unit.
%
From a utilitarian perspective, 
this bound is met even with just $k$ items. Suppose the
$i$th agent has an utility of $1-(k-1)\epsilon$
for the $i$th item, and $\epsilon$ for all other
items where $\epsilon$ is a small non-zero
constant. Note that the sum of the utilities
for any agent is normalized to 1 unit. 
The optimal utilitarian offline allocation
a social welfare of $k$ units
as $\epsilon$ goes to zero, whilst the expected 
utilitarian social welfare of the \mylike\
mechanism is just $1$ unit. 
\myqed
\myOmit{
For example, with 2 agents and general
utilities, the \mylike\ mechanism
is 2-competitive. That is, the expected egalitarian
or utilitarian social welfare
is at least 50\% of the optimal (offline) allocation. }

On the other hand, the \myblike\
mechanism is not competitive even with just 2 agents. 

\begin{mytheorem}
With general utilities and 2 agents, the 
\myblike\
mechanism is not $c$-competitive from an egalitarian
or utilitarian
perspective for any constant $c$.
\end{mytheorem}
\myproof
Consider the fair division of 4 items with the following utilities,
where $\epsilon>0$ is a small positive constant.

 \begin{tabular}{r c c c c}
 	& $a$ & $b$ & $c$ & $d$\\ \hline
 	Agent 1 & $\epsilon$ & $1-2\epsilon$ & $0$ & $\epsilon$\\
 	Agent 2 & $0$ & $\epsilon$ & $\epsilon$ & $1-2\epsilon$
 \end{tabular}

\noindent
Note that the sum of the utilities
for any agent is normalized to 1 unit. 
Then the optimal egalitarian (utilitarian) offline allocation 
gives items $1$ and $2$ to the first agent and items $3$ and $4$
to the second agent. 
This has an egalitarian (utilitarian) social welfare of 
$1-\epsilon$ unit ($2-\epsilon$ units). 
On the other hand, the \myblike\
mechanism results in an egalitarian (utilitarian) social 
welfare of just $2\epsilon$ ($4\epsilon$), allocating items
$1$ and $4$ to agent 1 and the other two items to agent 2. 
\myqed

Finally, when restricted to 0/1 utilities,
every allocation of the \mylike\ or \myblike\ mechanism
achieves the utilitarian
social welfare of the optimal 
offline allocation. The is because items only go to 
agents that value them.

\section{Price of anarchy}

The price of anarchy is closely related
to the competitive ratio but also takes
into account agents acting strategically. 
The price of anarchy 
measures how the efficiency 
of a decentralized system degrades due to selfish behavior of its agents
compared to imposing a centralized
solution based on sincere preferences \cite{poa}. 
From an egalitarian (utilitarian) 
perspective, the price of anarchy of an
online fair division mechanism is the ratio between
the optimal egalitarian (utilitarian) social welfare,
and the smallest egalitarian (utilitarian) social
welfare of any equilibrium strategy.
We consider \nPNEs (defined in Section~\ref{sec:welfare}).

\begin{mytheorem}
With $k$ agents, the
price of anarchy of the \mylike\ mechanism
is $k$ for egalitarian welfare,
and for utilitarian welfare it is at most $k$
and greater than $k-\epsilon$ for any $\epsilon>0$. 
\end{mytheorem}
\myproof
Note that we consider general utilities%
, and the \mylike\ mechanism is not strategy-proof in this case. 
Consider the equilibrium strategy
with least expected egalitarian (utilitarian) social welfare.
Suppose an agent bids for an item with non-zero
utility. The worst case is when every other agent 
bids against them. This gives an expected utility
which is $\frac{1}{k}$ of the sum of their utilities.
By comparison, the best case
is that they receive the sum of their 
utilities. 

From an egalitarian perspective,
this bound is achieved when
$k^2$ items are divided between
$k$ agents, the first agent has utility 1
for the first $k$ items, zero for the rest, and
every other agent has utility 1 for every item.
Then it is a dominant strategy for the first
agent to bid for the first $k$ items, and
for all other agents to bid for every item.
This gives an expected egalitarian social welfare
of 1, compared to the optimal egalitarian social
welfare of $k$ units. 

For the utilitarian case,
select $\epsilon'$ such that $0<\epsilon'<\frac{\epsilon}{k\cdot(k-1)}$. 
The bound is achieved
when $k$ items are divided between
$k$ agents, the $i$th agent has utility $1-(k-1)\epsilon'$
for the $i$th item and $\epsilon'$ for the rest. 
The dominant strategy is for every agent
to bid for every item. In this case, 
the optimal utilitarian social welfare is
$k\cdot (1-(k-1)\cdot \epsilon')>k\cdot (1-\frac{(k-1)\cdot \epsilon}{k\cdot (k-1)})=k-\epsilon$ whilst the expected utilitarian social welfare of the
\mylike\ mechanism is $1$.
\myqed

For the \myblike\ mechanism, we have the following lower bounds
on the price of anarchy.

\begin{mytheorem}
With 0/1 utilities and $k$ agents, the
price of anarchy of the \myblike\ mechanism from
an egalitarian perspective is at least $k$.
\end{mytheorem}
\myproof
Consider $k^2$ items being
divided between $k$ agents. The 
first agent has utility 1 for the first
$k$ items and 0 for all remaining items.
The other agents have utility 1 for
all items.
The optimal egalitarian offline allocation 
gives the first $k$ items to the first agent,
and $k$ of the other items to each of the other
agents. 
This has an egalitarian social welfare of $k$ units.
On the other hand, 
a dominant strategy with the \myblike\ mechanism
is sincerity. This gives an expected 
egalitarian social welfare of 1. 
\myqed

\begin{mytheorem}
	With general utilities and $k$ agents, the
	price of anarchy of the \myblike\ mechanism from
	a utilitarian perspective is greater than $k-\epsilon$,
	for any $\epsilon>0$.
\end{mytheorem}
\myproof
Consider an instance with $k$ items.
Select $\epsilon'$ such that $0<\epsilon'<\frac{\epsilon}{k\cdot(k-1)}$.
For each $i\in \{1,\dots,k\}$, agent $i$ has utility $1-(k-1)\cdot \epsilon'$ for item $i$ and
utility $\epsilon'$ for all other items.
In the \myblike\ mechanism, sincere play is the dominant strategy, allocating one item to each agent. The probability that  agent $i$ receives item $i$
is $\frac{k-1}{k}\cdot\frac{k-2}{k-1}\cdot\dots\frac{1}{k-i+1} = 1/k$.
Thus, the expected utilitarian welfare is $1-(k-1)\cdot \epsilon'+(k-1)\cdot \epsilon'=1$.
The optimal offline strategy simply allocates item $i$ to agent $i$, for an utilitarian welfare of $k\cdot (1-(k-1)\cdot \epsilon') > k\cdot (1-\frac{(k-1)\cdot \epsilon}{k\cdot (k-1)})=k-\epsilon$.
\myqed

Finally, with 0/1 utilities and either 
mechanism, it is a dominant strategy for agents
only to bid for (a subset of)
the items for which they have utility. 
Hence, both mechanisms achieve the
optimal utilitarian social welfare.
Thus, there is no price of anarchy from
an utilitarian perspective in these cases.

\myOmit{
\section{Non-randomized mechanisms}

Finally, we can derandomize these mechanisms.
For example, 
the {\em deterministic like} mechanism
is similar to the like mechanism except that to
we don't choose uniformly at random between agents 
that like an item 
but with a priority order that is rotated
each time it is used\footnote{For two agents,
it is simple to define a rotation procedure. For three or more,
it is more complex. For example, if the priority
order is initially agent 1 before 2 before 3 before 4, and we
have to tie-break between agents 2 to 4, then
we would allocate the item to agent 2. 
In addition, we rotate the position of the 3 tied
agents in the priority order
to give a new priority order of agent 1 before 3 before 4 before 2.}.
Similarly, the {\em deterministic biased like} mechanism
is similar to the biased like mechanism except that to
we don't choose uniformly at random between agents
but with a priority order that is rotated
each time it is used. 
Not surprisingly, derandomization increases the opportunity
for manipulation as agents can now exploit tie-breaking. 

\begin{mytheorem}
The deterministic like mechanism is not strategy-proof. 
\end{mytheorem}
\myproof
Suppose we are allocating items $a$ and $b$ between 2 agents,
agent 1 values both items, whilst
agent 2 only values item $b$.
The items are allocated in alphabetical order.
The initial priority order between agents
is agent 1 before 2. 
Suppose agents act sincerely and bid for all items
that they like. Then only agent 1 bids for the first
item $a$. In the second round, both agents bid for
item $b$. By the priority order, agent 1 is allocated
the item. 
By acting strategically, agent 2 can improve the
outcome. In particular, they can strategically bid
for item $a$ in the first round. Whilst agent 1 is
still allocated this item, agent 2 now gets priority
over agent 1 in the second round. Hence, agent 2 is 
allocated item $b$. 
\myqed

\begin{mytheorem}
The deterministic biased like mechanism is not strategy-proof. 
\end{mytheorem}
\myproof
Suppose we are allocating items $a$, $b$ and $c$ between 2 agents,
agent 1 values items $a$ and $c$, whilst
agent 2 values item $b$ and $c$.
The items are allocated in alphabetical order.
The initial priority order between agents
is agent 1 before 2. 
Suppose agents act sincerely and bid for all items
that they like. Then only agent 1 bids for the first
item $a$. In the second round, only agent 2 bids for
item $b$. Finally, in the third round, both agents bid
for item $c$ and, by the priority order, agent 1 is allocated
the item. 
By acting strategically, agent 2 can improve the
outcome. In particular, they can strategically bid
for item $a$ in the first round. Whilst agent 1 is
still allocated this item, agent 2 now gets priority
over agent 1 in the second round. Even if agent 1
bids for item $b$ in the second round, 
agent 2 is allocated the item and retains
the priority order. Hence, in the third round,
when both agents bid for item $c$, agent 2 is allocated
the item. This is a strict improvement
\myqed

\begin{mytheorem}
Even with 0/1 utilities, 2 agents and just 2 items, no
deterministic mechanism is $c$-competitive from an egalitarian
perspective for any constant $c$.
\end{mytheorem}
\myproof
We consider allocating two items in an online fashion.
Suppose both agents have utility 1 for the
first item. As the mechanism is deterministic,
we suppose without loss of generality that
the item is given to the first
agent. If this is not the case, we simply
remain the agents. Now suppose only the first agent has 
non-zero utility for the second item. Irrespective
of the allocation of the second item, the
egalitarian welfare of the final outcome is
0 as the second agent is not allocated the
one item that they value. On the other 
hand, the optimal (offline) allocation
assigns the first item to the second agent,
and the second item to the first agent. 
This has an egalitarian welfare of 1 unit. 
\myqed

This negative result supports why
we consider {\em randomized} mechanisms 
like the Like mechanism. 
}

\section{Experiments}

To determine the impact on social
welfare of these mechanisms 
and to determine if \myblike\ outperforms
\mylike\ in practice,
we ran some experiments.
We used a wide range of problem
instances: random 0/1 utilities,
random Borda utilities, 
correlated 0/1 and Borda utilities generated
with the P\'{o}lya-Eggenberger model,
as well as 0/1 and Borda utilities from PrefLib.org \cite{preflib}. 
For reasons of space, we report here just results
with random 0/1 utilities. 
We observed similar trends with the other
classes.

\begin{figure}[htb]
\centering
\includegraphics[width=0.5\columnwidth,clip=true,trim=19 0 44 25]{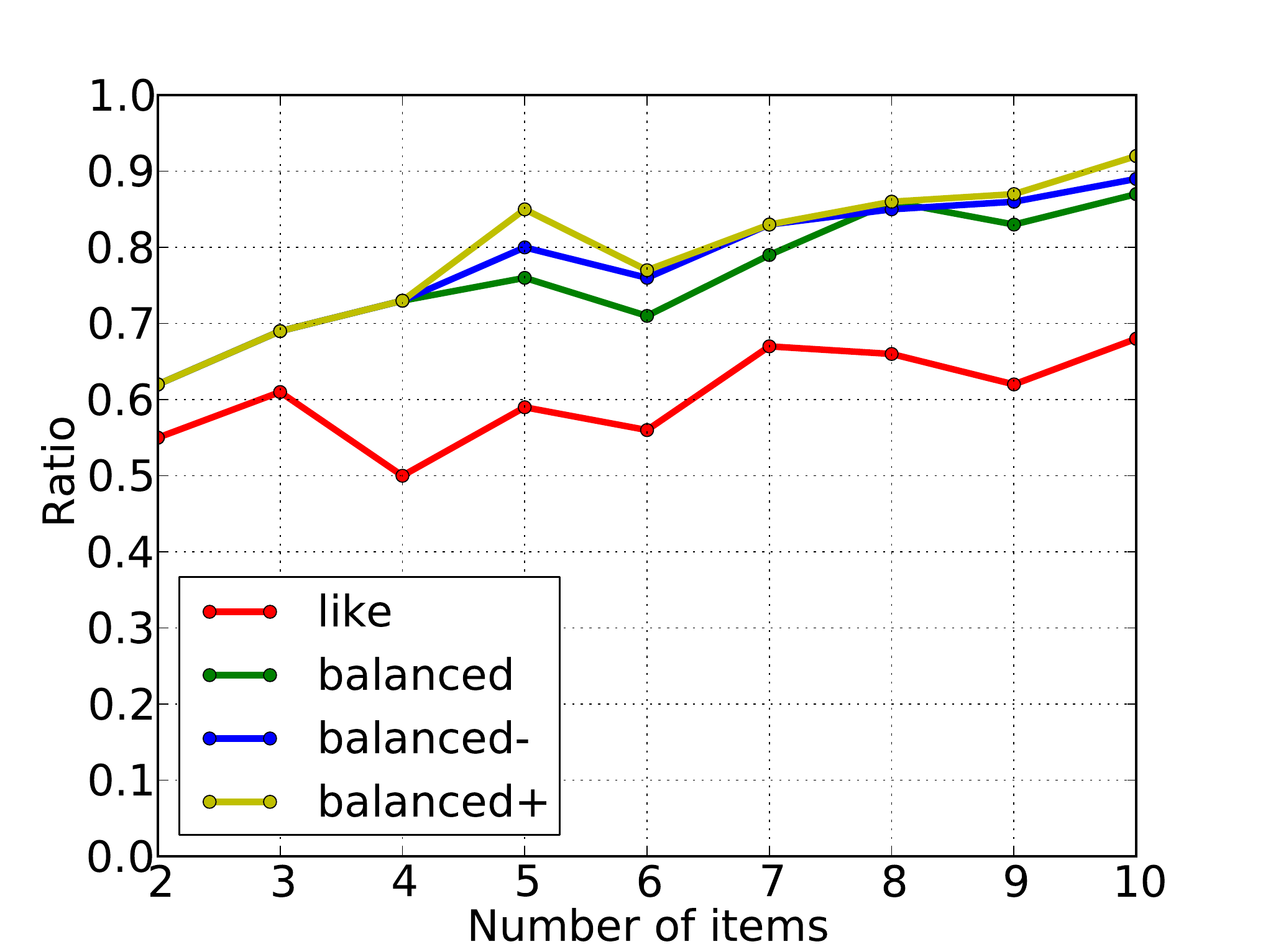}%
\includegraphics[width=0.5\columnwidth,clip=true,trim=19 0 44 25]{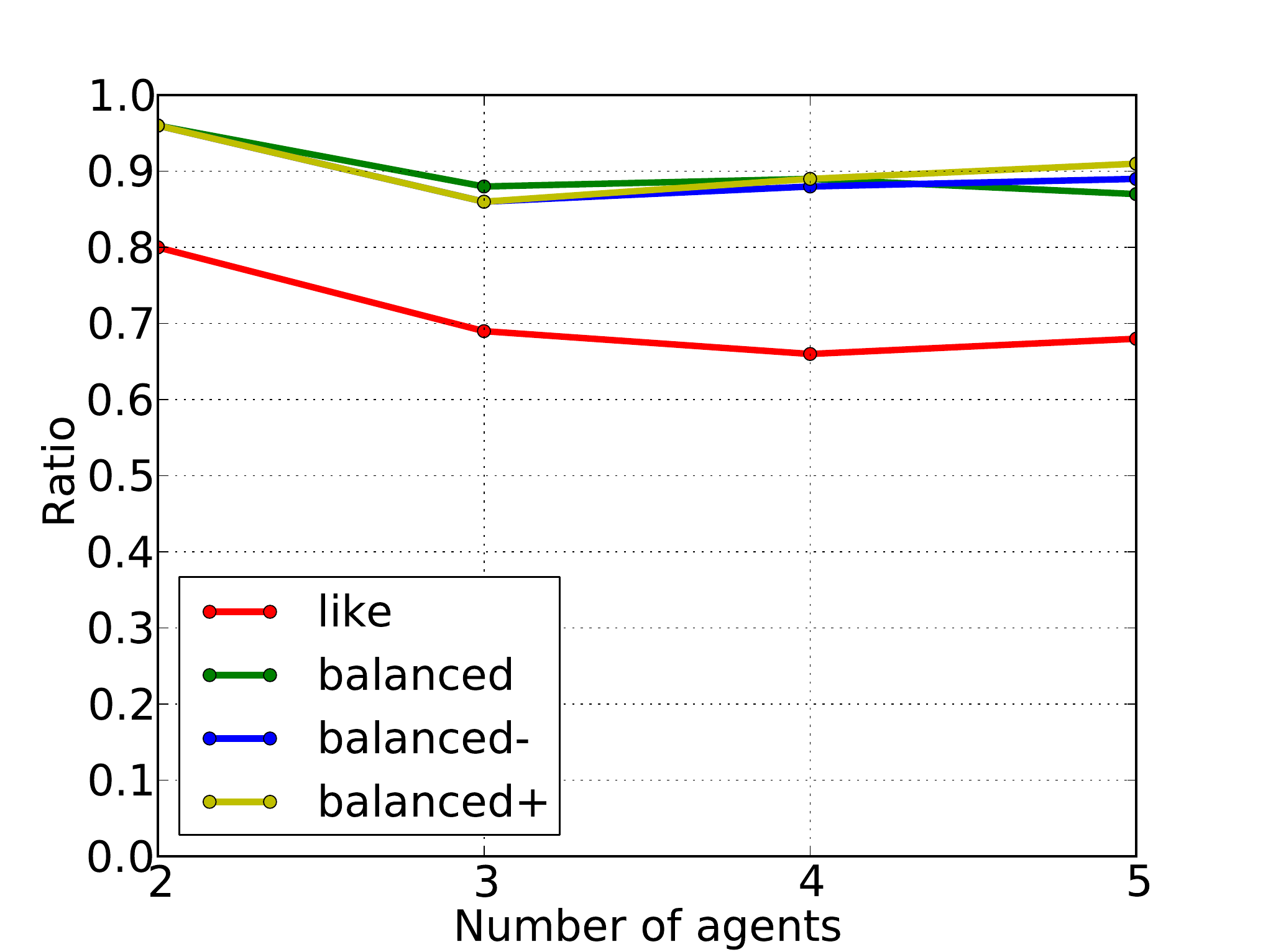}

\caption{Egalitarian price of anarchy, and competitive ratio of 
\myblike\ and \mylike\ mechanisms.
(left) varying items for 5 agents, (right) varying agents for 10 items.
}
\end{figure}

We varied the number of agents
from 2 to 5, and the number of items
from 2 to 10. We sampled 100 instances
at each data point, computing the optimal
(offline) allocation, and 
all \nPNEs\ by brute force. In Figure 1, we 
plot (1) the
competitive ratios (``like" and ``balanced"), (2) the prices of anarchy (``balanced-") and (3) the ratio
between the egalitarian welfare of the 
best \nPNE\ and the optimal
allocation (``balanced+"). As these are ratios,
we plot geometric means. Arithmetic
means are similar. We note that the \myblike\ mechanism (``balanced") improves the egalitarian welfare compared to the \mylike\ mechanism (``like") supposing sincere or strategic play of the agents. Indeed, strategic play of the agents often increases social welfare even in the worst case (``balanced-" compared to ``balanced"), though the effect is small. With Borda utilities, strategic play is less helpful and can result in lower social welfare. Nevertheless,
\myblike\ remained superior in all our experiments
to the \mylike\ mechanism. 

\section{Related work}

There is a large literature on the fair division
of divisible and indivisible goods. 
Almost all studies assume that all the goods
are present initially. There are, however,
a few exceptions. Walsh \shortcite{wadt11}
has proposed an online model of cake cutting.
However, in this model the agents arrive over
time (not the items), and the goods are 
divisible (not indivisible). Kash, Procaccia
and Shah \shortcite{kpsjair2014}
have proposed a related model in which
agents again arrive over time, but there are now
{\em multiple}, {\em homogeneous} {\em divisible} goods
(and not a single heterogeneous divisible good as in \cite{wadt11},
or multiple, heterogeneous indivisible goods as here). 
Bounded envy freeness is closely related to
the ``single-unit utility difference'' property that Budish, 
Che, Kojima and Milgrom \shortcite{bckmaec13}
prove can be achieved in {\em offline} fair division with any 
randomized allocation mechanism that is envy free ex ante.

The \mylike\ and \myblike\ mechanisms take an item-centric
view of allocation. They iterate over the items, allocating
them in turn to agents. By comparison,
there are agent-centric mechanisms like
the sequential allocation procedure which iterate over the
agents, allocating items to them in turn \cite{winwin}.
These mechanisms have attracted considerable attention
in the AI literature recently (e.g. 
\cite{indivisible,knwxaaai13,knwxaaai13}). 
As our matching problem is one-sided (agents have preferences
over items, but not vice-versa), we cannot
immediately map results from there to here. 
There are also randomized mechanisms like
random serial dictator \cite{rsd}, and the probabilistic 
serial mechanism \cite{probserial} which again take an agent-centric
view of allocation. It would be interesting future
work to consider how such agent-centric mechanisms 
could be modified to work with online fair division
problems. 

\myOmit{
\begin{table*}[htb]
{
\begin{center}
\begin{tabular}{|c|c|c|} \hline
 & \mylike\ mechanism & \myblike\ mechanism \\ \hline
strategy-proof & $\checkmark$ & $\times$, $\checkmark$ for $k$=2 \& 0/1 utilities   \\
envy free (ex ante) & $\checkmark$ & $\times$, $\checkmark$ for 0/1 utilities  \\
bound envy free (ex post) & $\times$ even for $k=2$ \& 0/1 utilities & $\times$, $\checkmark$ for 0/1 utilities \\
competitive (egalitarian or utilitarian)
 & $\checkmark$ & $\times$ even for $k$=2 \\
price of anarchy (egalitarian) & $k$ & $\geq k$ \\
price of anarchy (utilitarian) & $k$, 1 for 0/1 utilities & $\ge k$, 1 for 0/1 utilities \\ 
\hline      
\end{tabular}
\end{center}
}
\caption{Overview of results for $k$ agents}
\end{table*}
}

\section{Conclusions}

Motivated by our work with a local Food Bank charity,
we have studied a simple online model of fair division,
as well as two simple mechanisms for this problem. 
To help decide what mechanism to use in practice,
we have studied the axiomatic properties of these
mechanisms like strategy-proofness and envy-freeness.
In addition, 
we have undertaken a competitive analysis, and
computed their price of anarchy. 
A summary of our results is given in Table 1.

\begin{table}[tb]
{\scriptsize
\centering
\resizebox{\columnwidth}{!}{
\begin{tabular}{|c|c|c|} \hline
 & \mylike\ mechanism & \myblike\ mechanism \\ \hline
strategy-proof & $\checkmark$ & $\times$, $\checkmark$ for $k$=2 \& 0/1 utilities   \\
envy free (ex ante) & $\checkmark$ & $\times$, $\checkmark$ for 0/1 utilities  \\
bound envy free (ex post) & $\times$ even for & $\times$, $\checkmark$ for 0/1 utilities \\
& $k=2$ \& 0/1 utilities & \\
competitive 
 & $\checkmark$ & $\times$ even for $k$=2 \\
price of anarchy (e) & $k$ & $\geq k$ \\
price of anarchy (u) & $k$, 1 for 0/1 utilities & $\ge k$, 1 for 0/1 utilities \\ 
\hline      
\end{tabular}
}
}
\caption{Overview of results for $k$ agents. (e) = egalitarian, (u) = utilitarian. }
\end{table}

One possible take home message from
this table is that we might 
consider the \myblike\ mechanism if
the items can be packaged together
so that agents have similar utility
for all packages, and that we should
otherwise prefer the \mylike\ mechanism
when this is not possible. 
In future work, we plan to take into
account other important features of this
real world allocation problem. 
For example, as the charities have
different abilities to feed their clients,
we need a model of online fair division in
which the agents have different
entitlements.
Our 
mechanisms can be easily adapted to take
this feature into account.
\myOmit{For instance, in the
\mylike\ mechanism, we can simply bias the random choice of agent
by the size of their entitlement. }%
We will 
need to consider the impact this has on axiomatic
properties like strategy-proofness and fairness.
We will then be in a position to 
implement and field a mechanism for
online fair division in the field. 


\newpage

\bibliographystyle{named}
\bibliography{/Users/twalsh/Documents/biblio/a-z,/Users/twalsh/Documents/biblio/a-z2,/Users/twalsh/Documents/biblio/pub,/Users/twalsh/Documents/biblio/pub2}

\end{document}